\documentclass[A4,paper,12pt]{article}

\usepackage{amsmath}
\usepackage{amssymb}
\usepackage{graphicx}
\usepackage[cp1250]{inputenc}

\renewcommand{\[}{\begin{equation}}                               %
\renewcommand{\]}{\end{equation}}
\renewcommand{\d}{\mathsf{d}}
\newcommand{\pd}{\partial}
\newcommand{\R}{\mathbb{R}}

\newenvironment{proof}{\noindent\emph{Proof:}}{$\blacksquare$ \\ \smallskip}

\newtheorem{claim}{claim}
\newtheorem{theorem}[claim]{Theorem}
\newtheorem{remarks}[claim]{Remarks}
\newtheorem{proposition}[claim]{Proposition}

\begin{document}

\title{On the critical exponent in an isoperimetric inequality for chords}

\author{Pavel Exner$^{1,3}$, Martin Fraas$^{2,3}$, Evans M. Harrell II$^4$}
\date{\small 1) Nuclear Physics Institute, Czech Academy of Sciences,
25068 \v{R}e\v{z} near Prague, \\
2) Faculty of Mathematics and Physics, Charles University, V
Hole\v{s}ovi\v{c}k\'ach 2, 18040 Prague \\
3) Doppler Institute, Czech Technical University, B\v{r}ehov\'{a}
7, 11519 Prague, Czechia \\
4) School of Mathematics, Georgia Institute of Technology,
Atlanta, GA 30332, U.S.A. \\
\emph{e-mail: exner@ujf.cas.cz, fraas@ujf.cas.cz,
harrell@math.gatech.edu}}

\maketitle


{\small \noindent The problem of maximizing the $L^p$ norms of
chords connecting points on a closed curve separated by arclength
$u$ arises in electrostatic and quantum--mechanical problems.  It
is known that among all closed curves of fixed length, the unique
maximizing shape is the circle for $1 \le p \le 2$, but this is
not the case for sufficiently large values of $p$.  Here we
determine the critical value $p_c(u)$ of $p$ above which the
circle is not a local maximizer finding, in particular, that
$p_c(\frac12 L)=\frac52$.  This corrects a claim made in
\cite{EHL}. }

\vspace{3em}

\noindent
If $\Gamma(s)$ describes a planar
curve of parametrized by arclength $s$
and $L$ is its total length, then
$$
\left(\frac{1}{L}\int\limits_0^L{|\Gamma(s+u) -
\Gamma(s)|^p \d s}\right)
$$
describes the $L^p$-mean of the Euclidean length of the chords
connecting points separated by arclength $u$.  A reasonable geometric
question is to determine the shape that maximizes this quantity for
any given value of $p$.  Some physical phenomena have recently
been shown to have connections to this geometric question:
\begin{enumerate}
\item{What shape will a loop in $\R^3$ carrying a uniform electric
charge assume at equilibrium?  That is, what is the minimum of the
potential energy due to Coulomb repulsion?  For this problem see
\cite{ACF, JOH} and references therein. } \item{ What is the shape
of a loop $\Gamma$ of length $L$ that maximizes the ground-state
energy of a leaky quantum graph in the plane? That is, how can the
fundamental eigenvalue of the leaky-graph Hamiltonian
$-\Delta-\alpha\delta(x-\Gamma)$ acting in $L^2(\R^2)$ be
maximized?  This problem was considered in \cite{Ex, EHL} and
references therein. }
\end{enumerate}

In both of these problems it turns out that the solution reduces
to considering the $L^p$-means of chords, specifically to
establishing the validity of
\[ C_L^{p}(u):\quad c^p_\Gamma(u):=\int\limits_0^L|\Gamma(s+u) -
\Gamma(s)|^p \d s \leq \frac{L^{1+p}}{\pi^p} \sin^p \frac{\pi
u}{L}, \nonumber \]
with $p=1$ and $u \in (0,\frac12 L]$.  In other words, can it be
shown that the global maximizer is a planar circle of radius
$\frac{1}{2\pi}L$, which by an elementary calculation attains the
value on the right side? By a convexity argument it suffices to
prove the inequality for any larger value of $p$ to establish it
for smaller values.

The inequality $C_L^{p}(u)$ was established
for the first time over forty years ago by  L\"uk\H{o}
\cite{GL} for $p=2$. The same claim was demonstrated
more recently in
different ways in \cite{ACF, EHL}; see also a local proof in
\cite{Ex}.  It is natural to consider the maximal
value of $p$ for which the inequality holds. The best upper
estimate so far, $p\approx 3.15$, was obtained in \cite{EHL} by
investigating a stadium-shaped $\Gamma$.

Our aim here is to improve this result. Using the method of
\cite{Ex} we shall show that among all planar closed curves,
$c^p_\Gamma(u)$ is locally maximized
by a
circle if $p < \frac52$, and to find a local critical value of $p$
for ``shorter'' chords. Since the inequality in question has
obvious scaling properties, it is sufficient to consider the case
$L=2\pi$. We keep a general $L$ in the main claims for
the convenience
of the reader, but otherwise we will work with the
particular value $L=2\pi$.

Without loss of generality we may assume that the $\Gamma$ is a
$\mathcal{C}^2$-smooth curve, the validity of the result being
extended to less regular loops by continuity. Using the notation
of \cite[Sec.~5]{Ex} the quantity $c^p_\Gamma(u)$ can be cast into
the form
\[ \label{chordnorm}
c^p_\Gamma(u)=\int\limits_0^L \d s \left[\int\limits_s^{s+u} \d s'
\int\limits_s^{s+u} \d s'' \cos \left( \int\limits_{s'}^{s''}
\gamma(\tau) \d \tau \right) \right]^{p/2}, \nonumber
\]
where $\gamma:= \dot\Gamma_2 \ddot\Gamma_1- \dot\Gamma_1
\ddot\Gamma_2$ is the signed curvature of $\Gamma$. Recall that
the knowledge of $\gamma$ allows us reconstruct $\Gamma$ up to
Euclidean transformations by
 \begin{equation} \label{param}
 \Gamma(s)= \left( \int_0^s \cos\beta(t)\, \d t,
 \int_0^s \sin\beta(t)\, \d t \right)\,,
 \end{equation}
where $\beta(s):= \int_0^s \gamma(t)\, \d t$ is
the angle between the tangent vectors at
$t=s$ and the initial point, $t=0$.  We shall refer to this as the
{\it bending} of the arc.

Our aim is to compute the first and second G\^ateaux derivatives of
the map $\Gamma \mapsto c^{p}_\Gamma(u)$ at the circle, $\Gamma =
C$, and to demonstrate the claim by looking into their properties.
Consequently, we shall consider gentle deformations of a circle,
which can be characterized by variations of the curvature
\[ \label{var}
\gamma(s) = \frac{2 \pi}{L} + \varepsilon g(s), \]
where $g$ is a continuous $L$-periodic function and $\varepsilon$
is small in the sense that $\varepsilon\|g\|_\infty\ll1$. The
periodicity and continuity make it possible to express $g$ through
its Fourier series
\[
 g(s) = a_0 + \sum_{n=1}^{\infty} a_n \sin\left(\frac{2 \pi n
 s}{L}\right) + b_n \cos \left(\frac{2 \pi n s}{L}\right)
 \nonumber
 \]
with $\{a\},\,\{b\}\in\ell^2$. We are interested in closed curves
$\Gamma$, so we ask now how this property is
reflected in Fourier series.
\begin{proposition} \label{close}
The tangent to $\Gamma \in \mathcal{C}^2$ corresponding to
(\ref{var}) is periodic with period $L$ if and only if $a_0=0$.
Furthermore, $\Gamma(0)=\Gamma(L)+ \mathcal{O}(\varepsilon^3)$
provided that
\[a_1=b_1=0 \quad \mathrm{and} \quad
\sum_{n=2}^\infty \frac{b_n b_{n+1} + a_n a_{n+1}}{n(n+1)} =
\sum_{n=2}^\infty \frac{a_{n+1} b_n - b_{n+1} a_n}{n(n+1)} =0.\nonumber
\]
\end{proposition}

\begin{proof} As mentioned above, we may henceforth
set $L=2\pi$. In view of the definition of $\beta(s)$ it is
clear that the tangent vector is continuous if $\beta(L)=2\pi$. In
our case the bending function is
\[
\beta(s) = s + \varepsilon \int_0^s g(t) \d t =:
s + \varepsilon b(s), \nonumber 
\]
and the condition simplifies to $\int_0^{2\pi} g(t) \d t =0$ which
holds \emph{iff} $a_0=0$. In view of (\ref{param}) the fact that
$\Gamma$ is closed means
$$
\left(\int_0^{2\pi} \cos \beta(s)\, \d s ,\,\int_0^{2\pi} \sin
\beta(s)\, \d s \right) =(0,\,0).
$$
For the terms on the right side of the last equation we have the
expansion
\begin{eqnarray}
\cos \beta(s) = \left(1-\frac12 \varepsilon^2 b^2(s)\right)\, \cos
s - \varepsilon b(s) \sin s\,  + \mathcal{O}(\varepsilon^3), \nonumber \\
\sin \beta(s) = \left(1-\frac12 \varepsilon^2 b^2(s)\right)\, \sin
s  + \varepsilon b(s) \cos s\,  + \mathcal{O}(\varepsilon^3).
\nonumber
\end{eqnarray}
Up to the third order in $\varepsilon$ we get thus the conditions
\begin{eqnarray}
\int_0^{2\pi} b(s)\, \cos s  \, \d s
=\int_0^{2\pi} b(s)\, \sin s \, \d s = 0, \label{eq:pod1} \\
\int_0^{2\pi} b(s)^2\, \cos s \, \d s =\int_0^{2\pi} b(s)^2\, \sin
s \, \d s =0. \label{eq:pod2}
\end{eqnarray}
It is convenient to rewrite the Fourier series for the curvature
deformation in the complex form, $g(s) = \sum_{n\ne 0} c_n\,
e^{ins}$ where $c_{-n}=\bar c_{n}$ and for $n>0$ we have
$c_n=\frac12 (b_n-ia_n)$. For $b(s)$ this yields the following
series:
$$
b(s) =\sum_{n\ne 0}\, \frac{ic_n}{n}\, \left[1- e^{ins}\right].
$$
Using orthonormality of the trigonometric basis we see that the
condition (\ref{eq:pod1}) requires $a_1 = b_1 = 0$. On the other
hand, the remaining condition (\ref{eq:pod2}) means that the
integral $\int_0^L b(s)^2\, e^{ins}\, \d s $ must vanish; with the
help of the above series we can express it in the following way,
\begin{equation*}
-\sum_{n,\,m\ne 0,\pm 1} \frac{c_n c_m}{nm} \int_0^{2\pi} e^{is}
\left[1- e^{ins}\right] \left[1- e^{ims}\right]  \, \d s =
\sum_{n\ne 0,\pm 1 } \frac{c_n \bar
c_{n+1}}{n (n+1)}\,,
\end{equation*}
and taking the real and imaginary part we arrive at the
claimed identities for $\left\{a_n\right\}$
and $\left\{b_n\right\}$. \end{proof}

After this preliminary let us turn to our proper subject. The
G\^ateaux derivative of the functional (\ref{chordnorm}) in the
direction $g$ is
 \begin{multline} \mathsf{D}_g c_\Gamma^p(u) =
 \left.\frac{\pd c_\Gamma^p(u)}{\pd
 \varepsilon}\right|_{\varepsilon=0} \\
 = -\frac{p}{2} \left[4
 \sin^2 \frac{u}{2} \right]^{p/2-1}
 \int\limits_0^{2\pi} \d s \int\limits_s^{s+u} \d s'
 \int\limits_s^{s+u}\d s'' \sin \left( \int\limits_{s'}^{s''}
 \, \d t\right)\int\limits_{s'}^{s''} g(\tau)\, \d \tau
 \label{eq:PrvaDerivace}
\end{multline}
again for $L=2\pi$, and the second derivative is
\begin{multline}
  \mathsf{D}^2_g c_\Gamma^p(u) = \left.\frac{\pd^2 c_\Gamma^p(u)}{\pd
 \varepsilon^2}\right|_{\varepsilon=0} \\
 = \frac{p}{2}\left(\frac{p}{2}-1\right) \left[4
 \sin^2 \frac{u}{2} \right]^{p/2-2}
 \int\limits_0^{2\pi} \d s \left( \int\limits_s^{s+u} \d s'
 \int\limits_s^{s+u}\d s'' \sin (s''\!-\!s')
 \int\limits_{s'}^{s''} g(\tau)\, \d \tau \right)^{\!2} \\ -\frac{p}{2}
 \left[4 \sin^2 \frac{u}{2} \right]^{p/2-1}
 \int\limits_0^{2\pi} \d s \int\limits_s^{s+u} \d s' \int\limits_s^{s+u}\d s''
 \cos (s''\!-\!s')
\left( \int\limits_{s'}^{s''} g(\tau)\, \d \tau \right)^{\!2}.
\label{eq:Druhaderivace}
\end{multline}
Rearranging the integrals in (\ref{eq:PrvaDerivace}) we get
\begin{multline}
   \int\limits_0^{2\pi} \d s \int\limits_s^{s+u} \d s'
 \int\limits_s^{s+u}\d s'' \sin (s''-s')
 \int\limits_{s'}^{s''} g(\tau)\, \d \tau \\= \int\limits_0^{2\pi} \d \tau
 \int\limits_{\tau - u}^\tau \d s \int\limits_s^\tau \d s'
 \int\limits_\tau^{s+u} \d s''
 \sin (s''-s') \, g(\tau)\, \d \tau \\=
 \left(4 \sin^2 u +
 u \sin u\right)\int\limits_0^L g(\tau)\, \d
 \tau  = 0,
 \nonumber
\end{multline}
which shows that for every $p>0$ the circle is either an extremal
or a saddle point.
(There are no solutions to $4 \sin u = - u$ in $\left[-\pi, \pi\right]$.)
In the next step we analyze the second
derivative to distinguish in between these two cases. Not
surprisingly, the answer depends on the value of $u$.
Our main result reads

\begin{theorem} \label{main}
For a fixed arc length $u\in (0,\frac12 L]$ define
\[ p_{c}(u) := \frac{4 - \cos\left(\frac{2 \pi u}{L}\right)}
{1-\cos\left(\frac{2 \pi u}{L}\right)}, \label{eq:CritConst} \]
then we have the following alternative. For $p > p_{c}(u)$ the
circle is either a saddle point or a local minimum, while for $ p
< p_{c}(u) $ it is a local maximum of the map $\Gamma\mapsto
c_\Gamma^p(u)$.
\end{theorem}

Before passing to the proof let us make a pair of comments.
\begin{remarks}
\end{remarks}
\begin{enumerate}
\item
It will be seen from the proof that in
the critical case $p = p_c(u)$, the higher
order derivatives of $c_\Gamma^p(u)$ come into play.
We shall not address
the critical case here. \\ [.2ex]

\item

It is natural to expect and easy to verify that for $p>p_{c}$
circle is in fact a saddle point of the functional.
\end{enumerate}

\begin{proof} We put again $L=2\pi$ and  analyze the terms of
the second derivative (\ref{eq:Druhaderivace}) separately. By a
straightforward computation using orthonormality of the
trigonometric basis the iterated integral in the first term
equals
\[
 \sum_{n=2}^\infty \left[a_n^2 \mathsf{fs_1}(n,\,u,\,p) + b_n^2
 \mathsf{fc_1}(n,\,u,\,p) \right],
 \nonumber
 \]
 where
 \begin{equation*}
 \mathsf{fs_1}(n,\,u,\,p) = \mathsf{fc_1}(n,\,u,\,p):=
 \frac{16\pi}{n\!-\!n^3}\left(-2 n \cos \frac{n u}{2}
 \sin^2 \frac{u}{2} +
 \sin u \sin \frac{n u}{2} \right)^2.
 \nonumber 
\end{equation*}
In the second term we rearrange the integrals before using the
Fourier series,
\begin{multline}
\int\limits_s^{s+u} \d s' \int\limits_s^{s+u} \d
s''\int\limits_{s'}^{s''} \d \tau \int\limits_{s'}^{s''} \d \tau'
\cos (s''-s')\,g(\tau)g(\tau') \\= 2\int\limits_s^{s+u}\d \tau
\int\limits_\tau^{s+u} \d \tau' \int\limits_s^\tau \d s'
\int\limits_{\tau'}^{s+u} \d s'' \cos (s''-s')\, g(\tau)g(\tau')
\\=: \int\limits_s^{s+u}\d \tau \int\limits_\tau^{s+u} \d \tau' \,
g(\tau) g(\tau')\, \mathsf{Int}(s,\,\tau,\,\tau'). \nonumber
\end{multline}
Hence the full integral in the second term of
(\ref{eq:Druhaderivace}) equals
\begin{multline}
\int\limits_0^{2\pi} \d s \int\limits_s^{s+u}\d \tau
\int\limits_\tau^{s+u} \d \tau' \, g(\tau) g(\tau')\,
\mathsf{Int}(s,\,\tau,\,\tau') \\ = \int\limits_0^{2\pi} \d \tau
\int\limits_\tau^{\tau + u} \d \tau' \int\limits_{\tau'-u}^{\tau}
\d s\, g(\tau) g(\tau')\, \mathsf{Int}(s,\,\tau,\,\tau') =:
\int\limits_0^{2\pi} \d \tau \int\limits_\tau^{\tau + u} \d
\tau'\, \mathsf{Int_2}(\tau,\,\tau')\, g(\tau) g(\tau'), \nonumber
\end{multline}
where
 \begin{equation*}
  \mathsf{Int_2}(\tau,\,\tau') := 2(\tau'-\tau-u)\big(\cos
(\tau'-\tau) + \cos u \big) + 4\big(-\sin(\tau'-\tau) + \sin
u\big). \nonumber
\end{equation*}
Finally we use the Fourier series and obtain an expression for the
iterated integral in the second term
\[
\int\limits_0^{2\pi} \d s \int\limits_s^{s+u} \d s'
\int\limits_s^{s+u}\d s'' \cos (s''\!-\!s') \left(
\int\limits_{s'}^{s''} g(\tau) \d \tau \right)^{\!2}
=\sum_{n=2}^\infty \left[a_n^2 \mathsf{fs_2}(n,\,u,\,p) + b_n^2
 \mathsf{fc_2}(n,\,u,\,p) \right], \nonumber
\]
where
\begin{multline}
  \mathsf{fs_2}(n,\,u,\,p) = \mathsf{fc_2}(n,\,u,\,p)
  := \frac{\pi}{n-n^3} \left(-6 n^2 + 2n^4 -
  2(n^2-1)^2 \cos u \right. \\\left. +
  (n+1)^2\cos (n-1)u + (n-1)^2
  \cos (n+1)u \right).
  \nonumber
\end{multline}
Now we put it together and get the second derivative in the form
\[
\mathsf{D}^2_g c_\Gamma^p(u) = \sum_{n=2}^\infty\, (a_n^2 +
b_n^2)\, \frac{2^p \pi  \sin^{p-2}\left(\frac{ u}{2}\right)}
{8(n-n^3)^2}\, p\, T(n,\,u,\,p), \label{2Dseries}
\]
where
\begin{multline}
T(n,\,u,\,p) :=-\Big(2n^4 - 6n^2 -
  2(n^2-1)^2 \cos u +
  (n+1)^2\cos (n-1)u \\ + (n-1)^2
  \cos(n+1)u \Big) +
  2(p-2)\left(-2 n \cos\left(\frac{n
  u}{2}\right)\sin\left(\frac{u}{2}\right) + 2
  \cos\left(\frac{u}{2}\right) \sin\left(\frac{n
  u}{2}\right)\right)^2.
  \label{eq:tnup}
\end{multline}
Since $\sin(u/2)$ is positive for $u \in (0,\,\pi)$, the sign of
each term in the second derivative series (\ref{2Dseries}) is
determined by that of $T(n,\,u,\,p)$. The equation
\[
T(2,\,u,\,p) = -16\big(4-p + (p-1) \cos u \big)\sin^4
\left(\frac{u}{2} \right) \nonumber
\]
gives $T(2,\,u,\,p) > 0 $ for $p>p_c(u)$, proving the easier part
of the alternative, namely that for $p>p_c(u)$ the circle fails to
be a local maximum of the map $\Gamma\mapsto c_\Gamma^p(u)$.

It is easy to check that $T(n,\,u,\,p)$ is strictly increasing
as a function of
$p$. Hence to prove the other part of the theorem it is
sufficient to show that $T(n,\,u,\,p_{c}(u))$ is negative for $n
\geq 3$. To this aim we define
\[S(n,\,u) =
-(1-\cos u)\: T(n,\,u,\,p_c(u))\,; \nonumber \]
we next prove that this function is positive for $n \geq
3$.

Inserting the critical exponent $p_c(u)$ into (\ref{eq:tnup}) we
obtain
\begin{multline}
 S(n,\,u)= -4 -10 n^2 +2n^4 + 2(n^2-1) \big(-2(n^2-2)\cos u + n^2
 \cos^2 u \big) \\ + 4 \cos(nu) \big(1-n^2 + (2+n^2) \cos u \big) + 12 n
 \sin u \sin(nu),
 \nonumber
 \end{multline}
and using the inequality $ (a \sin x + b \cos x)^2 \leq a^2 + b^2 $ we
get the bound
\begin{multline}
S(n,\,u) \geq -4 -10 n^2 +2n^4 + 2(n^2-1) \big(-2(n^2-2)\cos u +
n^2 \cos^2 u \big) \\ -4 \sqrt{\big(1-n^2 + (2+n^2)\cos u \big)^2
+ 9 n^2 \sin^2 u}.
 \nonumber
 \end{multline}
Hence $S(n,\,u)$ is positive whenever
\[
-4 -10 n^2 +2n^4 + 2(n^2-1)\big(-2(n^2-2)\cos u + n^2
 \cos^2 u \big) > 0
 \label{eq:cond1}
\]
and
\begin{multline}
\Big(-4 -10 n^2 +2n^4 + 2(n^2-1) \big(-2(n^2-2)\cos u + n^2
 \cos^2 u \big) \Big)^2 \\ > 16 \Big( \big(1-n^2 + (2+n^2)\cos u \big)^2 + 9 n^2
 \sin^2 u\Big).
 \label{eq:cond2}
 \end{multline}
The first condition (\ref{eq:cond1}) is a quadratic equation
in $\cos u$, and a calculation shows that it is satisfied for
$\cos u < 1- \frac{6}{n^2} $. Using the notation $\cos u = x$, the second
condition (\ref{eq:cond2}) simplifies to
\[
4 n^2 (n^2-1)^2 (8 + n^2(x-1))(x-1)^3 >0, \nonumber
\]
which provides us with a slightly stronger condition,
\[
\cos u < 1- \frac{8}{n^2}\, \label{eq:ConPlus}.
\]
The vicinity of zero has to be
regarded separately to prove the positivity
of $S(n,\,u)$ on the interval complementary to (\ref{eq:ConPlus}) .
By a straightforward computation the Taylor expansion of
$S(n,\,u)$ around zero equals
\[
S(n,\,u) =\frac{n^2 u^8}{40}\left(-\frac{1}{9} + \frac{n^2}{4} -
\frac{n^4}{6} + \frac{n^6}{36} \right) + \frac{u^{10}}{10!}
R_{10}, \label{eq:TaylorExp}
\]
where
for $n \geq 3$ and $u$ in
the complement of (\ref{eq:ConPlus})
the $\mathcal{O}(u^{10})$ term is bounded from below by
\[
    R_{10} \geq - 136 n^{10}\,.
    \nonumber
\]
Comparing the reminder with the first term on the right-side of
(\ref{eq:TaylorExp}), we observe that $S(n,\,u)$ is positive
for
\[
 u^2 < \frac{1}{40}\left(-\frac{1}{9} + \frac{n^2}{4} -
\frac{n^4}{6} + \frac{n^6}{36} \right) \frac{10!}{136 n^8}\,
 \label{eq:OdhadZhora}.
\]
 Now we use the inequality $\cos u \leq 1
- 7/16 u^2$ for $u \in (0,\frac65)$  to compare the intervals
(\ref{eq:ConPlus}) and (\ref{eq:OdhadZhora}). By simple analysis
we find out that for $n \geq 4$,
\[
1-\frac{8}{n^2} \leq 1 - \frac{7}{16}
\frac{1}{40}\left(-\frac{1}{9} + \frac{n^2}{4} - \frac{n^4}{6} +
\frac{n^6}{36} \right) \frac{10!}{136 n^8}\,, \nonumber
\]
and hence in this case the union of the intervals covers
$(0,\,\pi)$, which proves  that $S(n,\,u) \geq 0$ holds for $n
\geq 4$.

In the case $n=3$ the positivity of $S(n,\,u)$ is easily
established, as the function $S(3,\,u)$ simplifies now to
\[
S(3,u) = 2 \left(2\sin \frac{u}{2}\right)^8. \nonumber
\]
Since $T(2,\,u,\,p) < 0 $ holds for $p<p_c(u)$ the theorem is
proven. \end{proof}

\begin{figure}
  \includegraphics{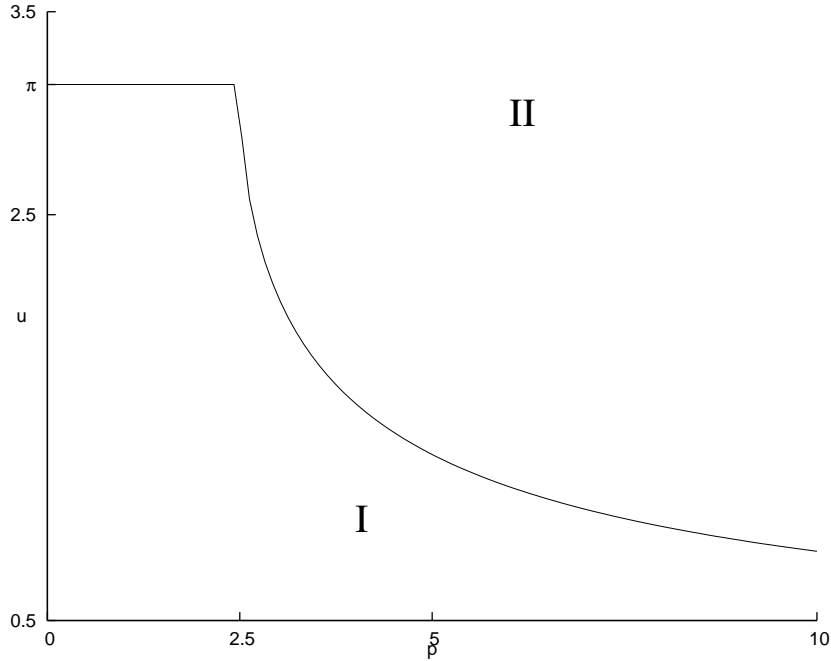} \\
  \caption{The relation between the critical exponent $p_c$
  and the arc length $u$. The mean-chord inequalities hold
  locally in the region I.}
  \label{fig:region}
\end{figure}

To visualize the result, in Figure~\ref{fig:region}
we plot
the relation between the critical exponent $p_c$ given by
(\ref{eq:CritConst}) and the arc length $u$.


A comment is due on the closure of the curve $\Gamma$. In
\cite{Ex} the local validity of the inequality for $p=2$ was
proved without this hypothesis. Here we used closure, but not to the
full power of Proposition~\ref{close}. We relied simply on the fact
that the Fourier coefficients vanish for $|n|\le 1$, which meant
that the endpoints $\Gamma(0)$ and $\Gamma(2\pi)$ meet within an
error of $\mathcal{O}(\varepsilon^2)$, not
$\mathcal{O}(\varepsilon^3)$.

Let us finally make one more remark, namely on a claim made in
Thm.~5.4 of \cite{EHL}. It was stated there that for a particular
class of deformations the circle remains a local maximizer for all
$p$, namely for those which, in the complex notation, have the
form $(1-\varepsilon) e^{is} + \Theta(\varepsilon,\,s)$, with the
assumption that for each $\varepsilon$, $\Theta(\varepsilon,\,s)$
is orthogonal to $e^{is}$ and $\Theta(\varepsilon,\,s)$ is
$\mathcal{C}^2$ smooth. In fact, the $\mathcal{C}^2$ assumption in
the variable $\varepsilon$ cannot occur. To see that, notice that
the condition $ \int |\dot\Gamma(s)|^2\, \d s = 2 \pi$ together
with orthogonality imply
$$ \int|\Theta_s(\varepsilon,\,s)|^2\, \d s
= 4 \pi \varepsilon -2 \pi \varepsilon^2,$$
where $\Theta_s:= \pd\Theta/\pd s$. Since $\Theta$ is
$\mathcal{C}^2$ by assumption, we may differentiate under the
integral sign to get
$$ 2 \mathrm{Re} \int \bar\Theta_s(\varepsilon,\,s) \frac{\pd
\Theta_s(\varepsilon,\,s)}{\pd \varepsilon}\: \d s = 4\pi - 4 \pi
\varepsilon\,;
$$
using the observation from \cite{EHL} that $\Theta(0,\,s)=0$ we
see that the left-hand side would tent to zero as $\varepsilon \to 0$
given the assumption that $\Theta$ is jointly $\mathcal{C}^2$,
while the right-hand one has the nonzero limit $4 \pi$.
To obtain smooth perturbations one should suppose, e.g.,
$\Gamma(\varepsilon,\,s) = (1-\varepsilon^2) e^{is} +
\Theta(\varepsilon,\,s) $, and this would necessitate an analysis
to second order in $\varepsilon$, as has been done in this
article.

\subsection*{Acknowledgments}

The research was supported in part by the Czech Academy of
Sciences and Ministry of Education, Youth and Sports within the
projects A100480501 and LC06002.

\end{document}